\documentclass[aps,amssymb,showpacs,twocolumn,superscriptaddress]{revtex4}
\usepackage{graphicx}

\def \auau  {Au + Au }
\def \rts   {$\sqrt{s_{NN}}$}
\def \jpsi  {$J/\psi$ }
\def\star{[STAR Collaboration]}
\def\phenix{[PHENIX Collaboration]}

\begin{document}

\title{\jpsi Production in Quark-Gluon
Plasma}

\author{Li Yan}
\affiliation{Physics Department, Tsinghua University, Beijing
100084, China}
\author{Pengfei Zhuang}
\affiliation{Physics Department, Tsinghua University, Beijing
100084, China}
\author{Nu Xu}
\affiliation{Nuclear Science Division, Lawrence Berkeley National
Laboratory, Berkeley, California 94720, USA}
\date{\today}

\begin{abstract}
We study \jpsi production at RHIC and LHC energies with both
initial production and regeneration. We solve the coupled set of
transport equation for the \jpsi distribution in phase space and
the hydrodynamic equation for evolution of quark-gluon plasma. At
RHIC, continuous regeneration is crucial for the $J/\psi$ momentum
distribution while the elliptic flow is still dominated by initial
production.  At LHC energy, almost all the initially created
$J/\psi$s are dissociated in the medium and regeneration dominates
the \jpsi properties.
\end{abstract}
\maketitle

The goal of high-energy nuclear collisions is to identify and
study the equation of state of the Quark-Gluon Plasma (QGP) which
is believed to exist at the early stage of our
universe~\cite{mueller041}. In \rts = 200 GeV \auau collisions,
the observations of jet-quenching~\cite{gyulassy03} and collective
flow have demonstrated the formation of hot and dense matter with
partonic collectivity~\cite{starwhitep,phenixwhitep}. Local
thermalization of the system created in heavy ion collisions is
yet to be tested.  Heavy flavors including charm and bottom quarks
are powerful tools~\cite{mueller04} for studying the early
collision dynamics because their masses are much larger than
possible excitation, temperature for example, of the system
created in collisions at RHIC. Recent results have shown that
heavy flavors are all produced in the initial
collisions~\cite{stare,phenixe}. Therefore, studying open charm
and charmonium production will yield important information on the
properties of QGP. The \jpsi is a particularly sensitive probe of
the early stages because its survial probability depends on the
environment. Lattice gauge theory calculations have indicated that
$J/\psi$s {\it do exist} above the critical
temperature~\cite{karsch,hatsuda}.

The $J/\psi$ suppression was first proposed as a direct signature
to identify the QGP formation twenty years ago~\cite{satz}.
Besides the normal suppression induced by nuclear absorption, the
$J/\psi$s initially produced by hard processes are anomalously
suppressed~\cite{blaizot,polleri,capella,bratkovskaya,hufner,zhu,sps}
by interactions in the hot medium. While charm quark production at
the SPS is expected to be small, there are more than 10 $c\bar{c}$
pairs produced in a central \auau collision at RHIC, and is
probably more than 200 pairs at the LHC~\cite{gavai}. These
uncorrelated charm quark pairs in the QGP can be recombined to
form $J/\psi$s. Obviously, regeneration will enhance the $J/\psi$
yield and alter its momentum spectra. Recently, the regeneration
approach for $J/\psi$ production at RHIC has been widely discussed
with different models, such as thermal production on the
hadronizaton hypersurface according to statistic
law~\cite{munzinger,gorenstein,rapp}, the coalescence
mechanism~\cite{greco} and the kinetic
model~\cite{grandchamp,thews} which considers continuous $J/\psi$
regeneration in a QGP.

The medium created in high-energy nuclear collisions evolves
dynamically. In order to extract information about the medium by
analyzing the \jpsi distributions, both the hot and dense medium
and the \jpsi production processes must be treated dynamically. In
this Letter, we treat continuous regeneration of the \jpsi in a
QGP self-consistently, including hydrodynamic evolution of the QGP
itself. Both regenerated and initially produced $J/\psi$s are
simultaneously taken into account. Including both processes is
important for RHIC since, unlike the collisions at the LHC, not
all the initially produced $J/\psi$s are destroyed.

The massive $J/\psi$s are unlikely fully thermalized with the
medium. Thus their phase space distribution should be governed by
transport rather than the kinetic equation of
Refs.~\cite{grandchamp,thews}.  To comprehensively treat the
$J/\psi$ distribution, we determine the $J/\psi$ transport
equation including both initial production and anomalous
suppression as well as regeneration. The transport equation is
then solved together with the hydrodynamic equation which
characterizes the space-time evolution of the QGP.  In order to
compare with the experimental measurements, we calculate the
average transverse momentum squared, $\left<p^2_t\right>$,
elliptic flow, $v_2$, and nuclear modification factor, $R_{AA}$,
for $J/\psi$ at both RHIC and LHC.

In p + p collisions, about 30-40\% of final state $J/\psi$s are
from the feed-down~\cite{zoccoli} of $\psi'$ and $\chi_c$ . In
order to simplify the numerical calculation, we neglect the
$\psi'$ contribution and assume 40\% of the final state $J/\psi$'s
come from $\chi_c$~\cite{zhu}. Since the $\Psi$ (=$J/\psi,\chi_c$)
is heavy, we use a classical Boltzmann-type transport equation to
describe its evolution. The distribution function $f_\Psi({\bf
p}_t,{\bf x}_t,\tau|{\bf b})$ in the central rapidity region and
in the transverse phase space (${\bf p}_t,{\bf x}_t$) at fixed
impact parameter ${\bf b}$ is controlled by the equation
\begin{equation}
\label{trans}
\partial f_\Psi/\partial \tau +{\bf
v}_\Psi\cdot{\bf \nabla}f_\Psi = -\alpha_\Psi f_\Psi +\beta_\Psi.
\end{equation}
The second term on the left-hand side arises from free-streaming
of $\Psi$ with transverse velocity ${\bf v}_\Psi = {\bf
p}_t/\sqrt{{\bf p}_t^2+m_\Psi ^2}$, which leads to the ``leakage''
effect and is needed to explain the averaged transverse momentum
squared at the SPS~\cite{hufner}. The anomalous suppression and
regeneration mechanisms are reflected in the loss term
$\alpha_\Psi$ and gain term $\beta_\Psi$, respectively.

Suppose the medium locally equilibrates at time $\tau_0$, when
nuclear absorption of the initially produced $J/\psi$s has ceased.
Absorption effects can be included in the initial
distribution~\cite{hufner,zhu}, $f_\Psi({\bf p}_t,{\bf
x}_t,\tau_0|{\bf b})$, of the transport equation. To determine the
initial condition at RHIC, we take $\tau_0=0.6$ fm, the initial
\jpsi production cross section $\sigma_{pp}^\Psi=2.61 \pm 0.20\
\mu$b~\cite{adler} and nuclear absorption cross section
$\sigma_{abs}=3$ mb~\cite{vogt}. We also assume that
$a_{gN}=0.076$ GeV$^2$/fm and $\left<p_t^2\right>_{pp}=4.31$
GeV$^2$~\cite{adler} to describe the initial $p_t$ broadening due
to multiple gluon scattering.

When the loss and gain terms $\alpha_\Psi$ and $\beta_\Psi$ are
known, the transport equation (\ref{trans}) can be solved
analytically with the result
\begin{eqnarray}
\label{distr}
f_\Psi({\bf p}_t,{\bf x}_t,\tau|{\bf b}) &=&
f_\Psi({\bf
p}_t,{\bf x}_t-{\bf v}_\Psi(\tau-\tau_0),\tau_0|{\bf b})\\
&&\times\ e^{-\int_{\tau_0} ^\tau d\tau' \alpha_\Psi({\bf
p}_t,{\bf x}_t-{\bf
v}_\Psi(\tau-\tau'),\tau'|{\bf b})}\nonumber\\
&&+\int_{\tau_0} ^\tau d\tau' \beta_\Psi({\bf p}_t,{\bf x}_t-{\bf
v}_\Psi(\tau-\tau'),\tau'|{\bf b})\nonumber\\
&&\times\ e^{-\int_{\tau'}^\tau d\tau''\alpha_\Psi({\bf p}_t,{\bf
x}_t-{\bf v}_\Psi(\tau-\tau''),\tau''|{\bf b})}.\nonumber
\end{eqnarray}
The first and second terms on the right-hand side indicate the
contributions from the initial production and continuous
regeneration, respectively. Both suffer anomalous suppression. The
coordinate shift ${\bf x}_t\to{\bf x}_t-{\bf v}_\Psi\Delta\tau$
reflects the leakage effect during the time period $\Delta\tau$.

We determine now the \jpsi suppression and regeneration in a QGP. We
consider only the gluon dissociation process, $g+\Psi\rightarrow c+\bar c$,
for the loss term~\cite{zhu}. Its inverse process is the gain
term. Then $\alpha_\Psi$ and $\beta_\Psi$ are
\begin{eqnarray}
\label{lg} \alpha_\Psi({\bf p}_t,{\bf x}_t,\tau|{\bf b}) &=&
{1\over 2E_\Psi}\int{d^3{\bf p}_g\over (2\pi)^3
2E_g}W_{g\Psi}^{c\bar
c}(s)f_g({\bf p}_g,{\bf x}_t,\tau)\nonumber\\
&&\times\ \Theta\left(T({\bf x}_t,\tau|{\bf
b})-T_c\right),\nonumber\\
\beta_\Psi({\bf p}_t,{\bf x}_t,\tau|{\bf b}) &=& {1\over
2E_\Psi}\int{d^3{\bf p}_g\over (2\pi)^3 2E_g}{d^3{\bf p}_c\over
(2\pi)^3 2E_c}{d^3{\bf p}_{\bar c}\over (2\pi)^3 2E_{\bar
c}}\nonumber\\
&&\times W_{c\bar
c}^{g\Psi}(s)f_c({\bf p}_c,{\bf x}_t,\tau|{\bf b})f_{\bar c}({\bf p}_{\bar c},{\bf x}_t,\tau|{\bf b})\nonumber\\
&&\times (2\pi)^4\delta^{(4)}(p+p_g-p_c-p_{\bar c})\nonumber\\
&&\times \Theta\left(T\left({\bf x}_t,\tau|{\bf
b}\right)-T_c\right),
\end{eqnarray}
where $E_g, E_\Psi, E_c$ and $E_{\bar c}$ are the gluon,
charmonium, $c$ and $\bar{c}$ energies. The step function $\Theta$
ensures that anomalous suppression and regeneration occur only in
the QGP phase. The gluon thermal distribution is $f_g$.
$W_{g\Psi}^{c\bar c}(s)$~\cite{cross} is transition probability of
the gluon dissociation as a function of $s=(p+p_g)^2$ calculated
in pQCD~\cite{cross}, and $W_{c\bar c}^{g\Psi}$ is $c$ and $\bar
c$ recombination transition probability. Note that $W_{c\bar
c}^{g\Psi}$ can be obtained from $W_{g\Psi}^{c\bar c}$ using
detailed balance.  In numerical calculations, we take $m_c=1.87$
GeV, including in-medium effects~\cite{grandchamp,zhu},
$m_{J/\Psi}=3.1$ GeV, $m_{\chi_c}=3.51$ GeV and critical
temperature $T_c = 165$ MeV.

Unlike gluons that are constituents of a QGP, charm quarks are
heavy and may not be thermalized in the QGP. In principle, a
self-consistent treatment of charm quark motion in the QGP should
be described by a transport equation similar to Eq.~(\ref{trans}).
For simplicity, we consider the $c$ and $\bar{c}$ distribution
functions, $f_c$ and $f_{\bar c}$, in two extreme scenarios. In
the weak interaction limit, the charm quarks in QGP are assumed to
keep their original momentum distribution, $g_c({\bf q})$,
calculated in pQCD~\cite{thews} and their initial space
distribution determined by nuclear geometry,
\begin{equation}
\label{pdis} f_{c,\bar c}({\bf q}, {\bf x_t}|{\bf
b})=\sigma_{pp}^{c\bar{c}}T_{A}({\bf x_t})T_{B}({\bf x_t}-{\bf
b})g_{c}({\bf q}),
\end{equation}
where $\sigma_{pp}^{c\bar{c}}=622 \pm 57\ \mu$b~\cite{cross2} is the
charm production cross section at RHIC. In the opposite limit, when
the $c$ and $\bar c$ are strongly correlated to the medium, they are
thermalized and distributed statistically. Both the pQCD and thermal
distributions are normalized to the initial charm quark number.

Since the hadronic phase occurs later in the evolution of heavy ion
collisions when the density of the system is lower compared to the
early hot and dense period, we have neglected the hadronic
dissociation. The suppression and regeneration region controlled by
the step function in the loss and gain terms and the gluon and
thermal charm quark distributions are determined by the QGP
evolution. With the Hubble-like longitudinal expansion and boost
invariant initial condition, the transverse hydrodynamical
equations~\cite{zhu} can be solved numerically to determine the
evolution of temperature and transverse fluid velocity.

\begin{figure}[h]
\centering
\includegraphics[totalheight=2.1in]{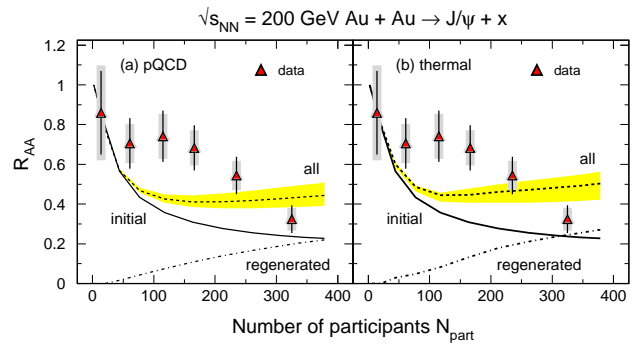}
\vspace{-1.15cm} \caption{The nuclear modification factor $R_{AA}$
as a function of participant number $N_{part}$ in pQCD (a) and
thermal (b) charm quark distributions. The solid and dot-dashed
curves are the calculations with only initial production and only
regeneration. The bands are the full result, including the
uncertainty in initial charm quark and $\Psi$ production. The data
are from PHENIX~\cite{phenix}.} \label{fig1}
\end{figure}

The final state $J/\psi$ distributions at fixed centrality can be
found by integrating the distribution function, Eq.~(\ref{distr}),
over ${\bf p}_t$ and ${\bf x}_t$ up to $\tau\rightarrow \infty$.
We first examine the mid-rapidity nuclear modification factor
$R_{AA}$. The numerical result, as a function of participant
number, $N_{part}$, is compared to the mid-rapidity RHIC data in
Fig.~\ref{fig1}. Plots (a) and (b) correspond to the pQCD and
thermal charm quark distributions, respectively. The bands are
theoretical results including the uncertainty in the initial charm
quark and $\Psi$ production cross sections $\sigma_{pp}^{c\bar c}$
and $\sigma_{pp}^\Psi$. The solid and dot-dashed curves indicate
calculations with only initial production (without regeneration,
$\beta_\Psi=0$) and only regeneration (without initial production,
$\sigma_{abs}=\infty$). Almost all the initial 40\% of $J/\psi$s
from $\chi_c$ decay is lost in semi-central collisions
~\cite{karsch06}. In central collisions, only the directly
produced $J/\psi$s suffer anomalous suppression. When nuclear
absorption effect is reduced by decreasing the initial time
$\tau_0$ or the absorption cross section $\sigma_{abs}$, the
initial $J/\psi$ yield is less suppressed. Due to the strong
regeneration in central collisions, the full $J/\psi$ yield is no
longer a monotonically decreasing function of centrality.  The
data seem to show a flat region at $50\le N_{part}\le150$. This
feature can not be reproduced in the present model calculations.

\begin{figure}[h]
\centering
\includegraphics[totalheight=2.1in]{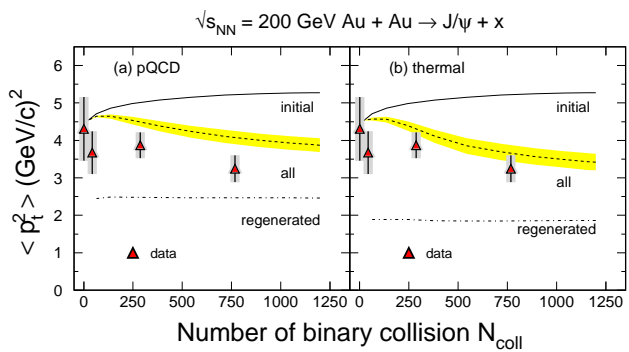}
\vspace{-1.15cm} \caption{The $J/\psi$ averaged transverse
momentum square as a function of the number of binary collision,
$N_{coll}$. The calculations are the same as in Fig.~\ref{fig1}.
The PHENIX~\cite{phenix} data are shown.} \label{fig2}
\end{figure}

The momentum spectra are expected to be more sensitive to the
production mechanism than the integrated yields. We show the
$J/\psi$ average squared transverse momentum, $\left<
p^2_t\right>$, as a function of centrality in Fig.~\ref{fig2}. It
is well known that the multiple gluon scattering in the initial
state leads to transverse momentum broadening. The leakage effect
due to the anomalous suppression also results in $p_t$ broadening
in central collisions~\cite{hufner}. These two broadening effects
on the initial $J/\psi$s are reflected in the solid curves in
Fig.~\ref{fig2}. For the regenerated $J/\psi$s, the charm quarks
in the pQCD scenario have no scattering in the initial state and
in the QGP, and they satisfy the statistic distribution in the
thermal scenario. The $\left<p_t^2\right>$ of the regenerated
$J/\psi$s in these two limits looks independent of centrality, see
dot-dashed curves in Fig.~\ref{fig2}. Since the high momentum
charm quarks lose energy during thermalization, the average
squared momentum ($\sim 2$ GeV$^2$) in the thermal scenario is
smaller than the one ($\sim 2.5$ GeV$^2$) in the pQCD scenario.
The collective flow develops with time, the $\left<p_t^2\right>$
for the sudden-produced $J/\psi$s on the hadronization
hypersurface~\cite{munzinger,gorenstein,grandchamp,greco} of QGP
is larger than our result for continuously regenerated $J/\psi$s
in the whole volume of QGP. Because the initial production
decreases with centrality, due to absorption and anomalous
suppression, and regeneration increases with centrality, shown in
Fig.~\ref{fig1}, the $\left<p_t^2\right>$ starts at the initial
result and approaches the regeneration result as centrality
increases.

\begin{figure}[h]
\centering
\includegraphics[totalheight=2.5in]{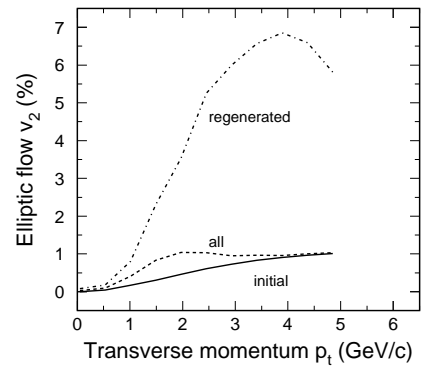}
\vspace{-1.0cm} \caption{The $J/\psi$ elliptic flow in $\sqrt{
s_{NN}} = 200$ GeV Au + Au collisions at impact parameter $b=7.8$
fm. The calculations are the same as in Fig.~\ref{fig1}.}
\label{fig3}
\end{figure}

The $J/\psi$ elliptic flow, $v_2$, is a useful tool for studying
charm production mechanism in nuclear collisions. When no
regeneration is considered, the leakage effect gives the lower
limit of $J/\psi$ $v_2$~\cite{zhu}, the solid curve in
Fig.~\ref{fig3}. In the pQCD scenario, charm quarks do not
interact with the medium and the regenerated $J/\psi$s do not
carry any collective property of the system, so we only discuss
$v_2$ in the thermal scenario. If the regeneration occurs only at
hadronization of the QGP, the well-developed azimuthal anisotropic
flow leads to a large $J/\psi\ v_2$, see discussions in
Ref.~\cite{greco}. However, for the continuous regeneration in the
QGP volume the average elliptic flow becomes much smaller since
the asymmetric flow is very small in the early stage. At impact
parameter $b=7.8$ fm, the continuously regenerated $J/\psi\ v_2$,
the dot-dashed line in Fig.~\ref{fig3}, is only one half of the
suddenly regenerated $J/\psi\ v_2$ in the coalescence
model~\cite{greco}. To check the difference between the continuous
and sudden regeneration mechanisms, we force the regeneration to
occur in a thin spherical shell by adjusting the regeneration
temperature region to $T_c\leq T\leq T_c+\delta T$. Indeed, the
regenerated $v_2$ approaches the value of Ref.~\cite{greco} for
$\delta T\rightarrow 0$. Since the $J/\psi$ yield in semi-central
collisions is still dominated by the initial production, see
Fig.~\ref{fig1}, there is no sizable difference in $v_2$ between
the total (dashed line) and initial-produced $J/\psi$s (solid
line), see Fig.~\ref{fig3}.

\begin{figure}[h]
\centering
\includegraphics[totalheight=2.5in]{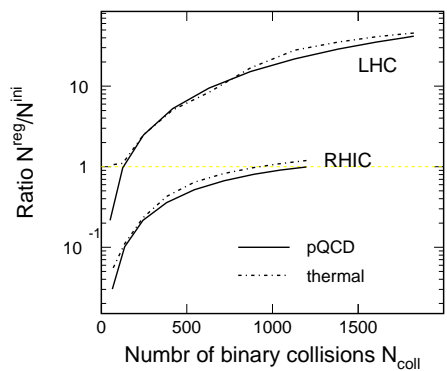}
\vspace{-1.0cm} \caption{The ratio of regenerated to initially
produced $J/\psi$ yields at RHIC and LHC energies in pQCD and
thermal scenarios. At the LHC, we choose $\sigma_{pp}^{\Psi}=18.9\
\mu$b~\cite{lhccross1}, $\sigma_{pp}^{c\bar c}=5740\
\mu$b~\cite{lhccross2}, $\tau_0=0.3$ fm and initial temperature
$680$ MeV.} \label{fig4}
\end{figure}

What is the case for nuclear collisions at the LHC? Due to the
extremely high center of mass energy, the QGP formed at LHC energy
will have much higher temperatures, longer lifetimes and larger
sizes. Most of the initial $J/\psi$s will be suppressed by
stronger gluon dissociation in QGP. On the other hand, more
charmonia will be created by the regeneration. The ratio
$N^{reg}/N^{ini}$ of regenerated to initial-produced $J/\psi$s is
shown as a function of centrality in Fig.~\ref{fig4}. While at
RHIC, the ratio can reach unity only in the most central
collisions, it is much larger than one at almost any centrality
bin at the LHC. Therefore, regeneration will dominate the observed
$J/\psi$. As a consequence, the \jpsi $R_{AA}$, $v_2$ and $\langle
p_t^2 \rangle$ at the LHC will follow the regeneration
calculations in Figs.~\ref{fig1},~\ref{fig2} and~\ref{fig3}.
\vspace{-1.0cm}

In summary, by constructing and solving the transport equation for
charmoniums together with hydrodynamic evolution of the QGP, we
presented a self-consistent way to calculate the $J/\psi$
transverse momentum distributions. Unlike the cases at the SPS,
where initial production is almost the only $J/\psi$ source and at
the LHC, where the regeneration dominates the observed $J/\psi$s,
both initial production and regeneration are important at RHIC.
The average squared transverse momentum with thermalized charm
quarks fits the RHIC data reasonably well. In contrast to previous
calculations~\cite{greco} with only sudden regeneration at
hadronization, the dominant initial production in semi-central
collisions and the continuous regeneration in the whole QGP volume
lead to a rather small $J/\psi$ elliptic flow at RHIC.

\vspace{0.1in} \noindent {\underline{Acknowledgments:} We are
grateful to R. Rapp and X. Zhu for the helpful discussions. We
thank Drs. R. Vogt and B. Mohanty for proof reading the
manuscript. P.Z.  thanks A. Kostyuk and H.St\"ocker for the useful
discussions and the Frankfurt Institute for Advanced Studies for
its financial support in the beginning of the work. The work is
supported in part by the Chinese grants No. NSFC10575058,
10425810, 10435080 and US DOE Contract No. DE-AC03-76SF00098.

\end{document}